\documentstyle[preprint,aps,epsf,floats]{revtex}
\begin{document}
\tighten

\def\bfl{{\bbox \ell}}
\def\bull{\vrule height .9ex width .8ex depth -.1ex}
\def\MeV{{\rm MeV}}
\def\GeV{{\rm GeV}}
\def\Tr{{\rm Tr\,}}
\def\nrcpt{NR\raise.4ex\hbox{$\chi$}PT\ }
\def\ket#1{\vert#1\rangle}
\def\bra#1{\langle#1\vert}
\def\ltap{\ \raise.3ex\hbox{$<$\kern-.75em\lower1ex\hbox{$\sim$}}\ }
\def\gtap{\ \raise.3ex\hbox{$>$\kern-.75em\lower1ex\hbox{$\sim$}}\ }
\newcommand{\gsim}{\raisebox{-0.7ex}{$\stackrel{\textstyle >}{\sim}$ }}
\newcommand{\lsim}{\raisebox{-0.7ex}{$\stackrel{\textstyle <}{\sim}$ }}

\def\Journal#1#2#3#4{{#1} {\bf #2}, #3 (#4)}

\def\NCA{\em Nuovo Cimento}
\def\NIM{\em Nucl. Instrum. Methods}
\def\NIMA{{\em Nucl. Instrum. Methods} A}
\def\NPB{{\em Nucl. Phys.} B}
\def\NPA{{\em Nucl. Phys.} A}
\def\PLB{{\em Phys. Lett.}  B}
\def\PRL{\em Phys. Rev. Lett.}
\def\PRD{{\em Phys. Rev.} D}
\def\PRC{{\em Phys. Rev.} C}
\def\PRA{{\em Phys. Rev.} A}
\def\PR{{\em Phys. Rev.} }
\def\ZPC{{\em Z. Phys.} C}
\def\PREP{{\em Phys. Rep.}  }
\def\ANN{{\em Ann. Phys.} }
\def\SCI{{\em Science} }
\def\CJP{{\em Can. J. Phys.}}

\preprint{\vbox{
\hbox{INT99-71}
\hbox{NT@UW-99-51}
}}
\bigskip
\bigskip

\title{Parity Violation in $\gamma \overrightarrow p$ 
Compton Scattering
}
\author{Paulo F. Bedaque}
\address{Institute for Nuclear Theory, University of Washington,
Seattle, WA 98195-1550
\\ {\tt bedaque@phys.washington.edu}}
\author{Martin J. Savage}
\address{Department of Physics, University of Washington, Seattle, 
WA 98195-1560
\\
and\\
 Jefferson Lab., 12000 Jefferson Avenue, Newport News, 
 VA 23606.
  \\ {\tt savage@phys.washington.edu}}
\maketitle

\begin{abstract}
A measurement of parity-violating spin-dependent
$\gamma \overrightarrow p$ 
Compton scattering will provide a theoretically clean
determination of the parity-violating pion-nucleon
coupling constant $h_{\pi NN}^{(1)}$.
We calculate the leading parity-violating amplitude
arising from one-loop pion graphs in chiral perturbation theory.
An asymmetry of $\sim 5\times 10^{-8}$ is estimated for Compton
scattering of $100\ {\rm MeV}$ photons.
\end{abstract}

\vfill\eject

Precise experimental work performed during the past decades  
has provided a catalogue of parity-violating matrix elements
in both light and heavy nuclei.
Unfortunately, a complete theoretical understanding of these measurements has
proved elusive thus far.
In particular, from the measurements in light nuclei one would hope to be able
to extract the parity-violating isovector pion-nucleon coupling constant, 
$h_{\pi NN}^{(1)}$, that is expected to provide the dominant source of 
$\Delta I=1$ parity
violation in the nucleon-nucleon potential.
However, a unique 
value of $h_{\pi NN}^{(1)}$ consistent with all measurements 
has not been established 
and, in addition, the $^{18}F$ measurements~\cite{Fluorine} 
suggest that 
$h_{\pi NN}^{(1)}$ is much smaller than naive 
estimates~\cite{DDH}-\cite{SKweak}, or that there are significant cancellations
between leading and sub-leading interactions\cite{KSa}.
This discrepancy is probably due to the difficulty of the nuclear physics
component of the  calculations, as opposed  to a signal of new physics.
Thus, the question remains as to which measurement or set of measurements
will most reliably determine  $h_{\pi NN}^{(1)}$.
Two-nucleon observables would seem to have a distinct advantage over other
multi-nucleon systems as the deuteron is so loosely bound.
Recently, a  proposal\cite{npprop} has been made 
to precisely determine $h_{\pi NN}^{(1)}$ from the 
forward-backward asymmetry $A_\gamma$ in the radiative capture 
of polarized neutrons by protons,
$\overrightarrow n+p\rightarrow d+\gamma$.  
The current experimental limit  is
$A_\gamma = -(1.5\pm 4.8)\times 10^{-8}$~\cite{Aetal},
while the proposed  experiment expects to measure $A_{\gamma}$ with
a precision of $\pm 5 \times 10^{-9}$.
Theoretically, if $h_{\pi NN}^{(1)}$  is of its naively estimated
size then  $A_\gamma$ will be dominated by the 
$h_{\pi NN}^{(1)}$ coupling\cite{Dani}-\cite{KSSWpv}.
If, on the other hand, $h_{\pi NN}^{(1)}$ is much smaller than
estimated, the existing calculations of this asymmetry will be invalid.
An alternate determination may be possible at Bates\cite{sample}, 
by a precise measurement of
deuteron spin-dependent parity-violation in electron-deuteron scattering.
This process is not ideal due to contributions from direct $Z^0$-exchange
(and higher order interactions) between the electron and the deuteron.
Nonetheless, a constraint will be placed on $h_{\pi NN}^{(1)}$ from such 
a measurement\cite{SSana,KKa}.
An analogous measurement in scattering from nucleons does not provide the same 
constraint due to the much larger 
isovector coupling of the $Z^0$, which is absent
(at tree-level) in the deuteron.
In this work we show that a precise measurement of Compton 
scattering from polarized protons
(and neutrons) 
will allow for a theoretically clean extraction of
$h_{\pi NN}^{(1)}$, which contributes through one-loop pion
graphs.

The strong interactions of the pions and nucleons are described 
at leading order in 
heavy baryon chiral perturbation theory ($HB\chi PT$)\cite{HBCPT}
by 
\begin{eqnarray}
{\cal L}_{st} & = &  {f^2\over 8} Tr D_\mu \Sigma
D^\mu \Sigma^\dagger
+ \overline{N} \left( i v_\mu D^\mu \ +\ {{\bf D}^2\over 2 M_N}
\right)N 
+ 2 g_A \overline{N} S_\mu {\cal A}^\mu N
\ +\ \cdots
\ \ \ ,
\label{eq:lagst}
\end{eqnarray}
where $N$ is the isospin doublet of nucleon fields with four velocity $v$,
$M_N$ is the nucleon mass,
$S_\mu$ is the covariant spin-operator,
$g_A~\sim~+1.25$ is the axial coupling constant,
$f=132\ \MeV$ is the pion decay constant,
$D^\mu$ is the covariant derivative, 
and the  ellipses represent operators involving more insertions of the light
quark mass matrix, meson fields, and derivatives.  
The pion fields are contained
in a special unitary matrix,
\begin{equation}
\Sigma = \xi^2 = \exp {2i\Pi\over f},\qquad \Pi =
\left(\begin{array}{cc}
\pi^0/\sqrt{2} & \pi^+\\ \pi^- & -\pi^0/\sqrt{2}\end{array} \right)
\ \ \ ,
\label{eq:pions}
\end{equation}
and the axial vector meson field is
${\cal A}_\mu = \partial_\mu\Pi/f\ +\ ...$.
The lagrange density in eq.~(\ref{eq:lagst}) and 
the Wess-Zumino term,
give the leading contributions to the
parity-conserving
$\gamma \overrightarrow p$ Compton scattering amplitude $T^{pc}$, 
which has the form\cite{HHKKa} (in the center of momentum frame)
\begin{eqnarray}
T^{\rm pc} & = & 
\overline{N}\ \left[\ 
A_1\ {\bf \epsilon}\cdot {\bf\epsilon}^{\prime *}
\ +\  
A_2\ {\bf \epsilon}\cdot \hat {\bf k}^\prime
 {\bf\epsilon}^{\prime *} \cdot \hat{\bf k}
\ +\ 
i 2 A_3\ {\bf S}\cdot \left({\bf\epsilon}^{\prime *}\times  
{\bf \epsilon}\right)
\ +\ 
i 2 A_4 \  {\bf S}\cdot \left(\hat{\bf k}^\prime\times \hat{\bf k}\right)
 {\bf \epsilon}\cdot {\bf\epsilon}^{\prime *}
\right.
\nonumber\\
& & \left.
+  
i 2 A_5\   {\bf S}\cdot \left[ \left({\bf \epsilon}^{\prime *}\times \hat{\bf
      k}\right) {\bf \epsilon}\cdot  \hat{\bf k}^\prime
- \left({\bf \epsilon}\times \hat{\bf k}^\prime\right)
{\bf \epsilon}^{\prime *}\cdot  \hat{\bf k}\right]
\right.
\nonumber\\
& & \left.
+  
i 2 A_6 \  {\bf S}\cdot \left[ \left({\bf \epsilon}^{\prime *}\times \hat{\bf
      k}^\prime\right) {\bf \epsilon}\cdot  \hat{\bf k}^\prime
- \left({\bf \epsilon}\times \hat{\bf k}\right)
{\bf \epsilon}^{\prime *}\cdot  \hat{\bf k}\right]
\right] \ N
\ \ \ ,
\label{eq:ampst}
\end{eqnarray}
where ${\bf S}$ are the three-vector components of $S_\mu$, 
and $\hat {\bf k}, \hat {\bf  k}^\prime$ are unit vectors in the direction
of   ${\bf k},  {\bf  k}^\prime$ respectively.
The $A_i$ are functions of the photon energy $\omega$
and  scattering angle $\theta$, and can
be found in\cite{HHKKa,JKOa,KMBa}.  
They receive contributions from tree-level and one-loop pion graphs, 
as well as from the Wess-Zumino term.  
At leading order in the chiral expansion arises from the covariant derivative
term in eq.~(\ref{eq:lagst}), and gives
\begin{eqnarray}
A_1 & = & -{e^2\over M_N}
\ \ \ ,
\label{eq:lowest}
\end{eqnarray}
with the remaining $A_i$ vanishing.
Starting at next order in the chiral expansion 
the other $A_i$'s become non-zero.

\begin{figure}[t]
\centerline{\epsfxsize=4.0in \epsfbox{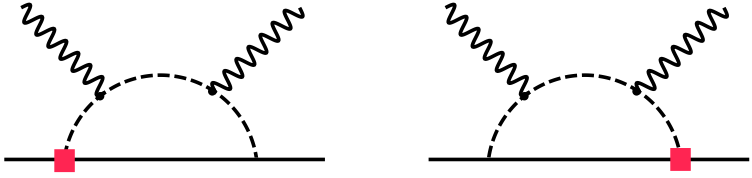}}
\noindent
\caption{\it 
  The leading order contribution to parity violation in 
$\gamma N$ Compton scattering.
  The solid square is the weak operator with 
  coefficient $h^{(1)}_{\pi NN}$.
  Wavy lines are photons, solid lines are nucleons, 
  and dashed lines are pions.
  We have not shown the crossed graphs.
  In addition we have not shown graphs with photons from the strong vertex,
  or insertion of the two-photon-pion vertex as they vanish in the 
  $v.A=0$ gauge.
  }
\label{fig:loops}
\vskip .2in
\end{figure}

The $\Delta I=1$ flavor-conserving-parity-violating 
interactions,
including $\gamma \overrightarrow N$ Compton scattering,
will be dominated by the 
lowest order operator in the chiral lagrangian 
\begin{eqnarray}
    {\cal L}^{\Delta I=1}_{\rm weak} & = &
-{ h_{\pi NN}^{(1)}\over\sqrt{2}}\ 
\varepsilon^{3\alpha\beta} \ \overline{N}\pi^\alpha\tau^\beta N\ + \cdots
\ =\ i\ h_{\pi NN}^{(1)}\ \pi^+\ p^\dagger n\ +\
  {\rm h.c.}\ + \cdots
\ \ \ ,
\label{eq:lagwk}
\end{eqnarray}
if its coefficient, $ h_{\pi NN}^{(1)}$,
is of natural size. 
The ellipses denote terms involving more pion fields required by chiral
invariance.
Explicit computation of the one-loop graphs shown in fig.~(\ref{fig:loops})
gives a parity-violating amplitude in the center-of-momentum frame of
the form
\begin{eqnarray}
T^{pv} \ =\  {e^2 g_A h_{\pi NN}^{(1)}\over 2\pi^2 f}\  
\overline{N}\ \tau_3\ & & 
\left[
{\cal F}_1(\omega,\theta)\ 
{\bf S}\cdot ({\bf k}+{\bf k}^\prime) 
\ {\bf \epsilon}\cdot {\bf \epsilon}^{\prime *} 
\ -\
{\cal F}_2(\omega,\theta)\ 
\left(
{\bf S}\cdot {\bf \epsilon}^{\prime *}\  {\bf k}^\prime\cdot {\bf \epsilon}
\ +\ 
{\bf S}\cdot {\bf \epsilon}\  {\bf k}\cdot {\bf \epsilon}^{\prime *}
\ \right)
\right.\nonumber\\
& & \left.
-  {\cal F}_3(\omega,\theta)\  {\bf k}\cdot {\bf \epsilon}^{\prime *}\ 
 {\bf k}^\prime\cdot {\bf \epsilon}\
{\bf S}\cdot ({\bf k}+{\bf k}^\prime) 
\right]\ N
\ \ \ ,
\label{eq:pvamp}
\end{eqnarray}
where $\tau_3$ is the isospin matrix.
For general kinematics the loop functions ${\cal F}_i$ are somewhat
complicated, and we present them as integrals over two Feynman parameters,
\begin{eqnarray}
{\cal F}_1(\omega,\theta) & = & \int_0^1dx\ \int_0^{1-x} dy\ (1-2 y)\ 
\left[{\cal I} (-1; x\omega, \tilde m^2) 
- {\cal I}(-1; -x\omega, \tilde m^2)\right]
\nonumber\\
{\cal F}_2(\omega,\theta) & = & 2\int_0^1dx\ \int_0^{1-x} dy\ y\ 
\left[{\cal I} (-1; x\omega, \tilde m^2) 
- {\cal I}(-1; -x\omega, \tilde m^2)\right]
\nonumber\\
{\cal F}_3 (\omega,\theta)& = &  2\int_0^1dx\ \int_0^{1-x} dy 
\ y\ (1-x-y)\ (2y-1)\ 
\left[ {\cal I}(-2; x\omega, \tilde m^2) 
- {\cal I}(-2; -x\omega, \tilde m^2)\right]
\nonumber\\
\tilde m^2 & = & 
m_\pi^2 + 2 y (1-x-y)\ \omega^2\ \left(1-\cos\theta\right)
\ \ \ ,
\label{eq:ints}
\end{eqnarray}
where the functions ${\cal I}(\alpha;b,c)$ are defined by Jenkins and Manohar
in \cite{HBCPT}
\begin{eqnarray}
{\cal I}(\alpha; b,c) & = & 
\int_0^\infty d\lambda\ \left(\lambda^2+2\lambda b  +c\right)^\alpha
\nonumber\\
{\cal I}(-1; \Delta, m^2) & = & - {1\over 2\sqrt{\Delta^2-m^2+i\epsilon}}
\log\left({\Delta-\sqrt{\Delta^2-m^2+i\epsilon}\over 
\Delta+\sqrt{\Delta^2-m^2+i\epsilon}}\right)
\nonumber\\
{\cal I}(-2; \Delta, m^2) & = & {1\over 2\left(\Delta^2-m^2+i\epsilon\right)}
\left({\Delta\over m^2} - {\cal I}(-1; \Delta, m^2)\right)
\ \ \ .
\label{eq:Idef}
\end{eqnarray}

For forward scattering, ${\bf k}={\bf k}^\prime$, 
the amplitude in eq.~(\ref{eq:pvamp}) collapses to 
$T^{pv}~\sim~\omega~{\bf \epsilon}~\cdot~{\bf \epsilon}^\prime~S~\cdot~k$, 
which is clearly parity violating.
As there is no non-derivative parity-violating coupling between the $\Delta$
and the nucleon, contributions from $\Delta$ intermediate states are suppressed
in the chiral expansion, unlike the situation for many other observables.
Therefore,
the one-loop contribution in eq.~(\ref{eq:pvamp})
is enhanced by two powers of the pion mass compared
to the naive size of local counterterms, whose size is set by $\Lambda_\chi$,
the scale of chiral symmetry breaking.  
However, there will be contributions at next order in the chiral expansion
that are suppressed by a single power of $m_\pi/\Lambda_\chi$ or
$\omega/\Lambda_\chi$, compared
to the contribution in eq.~(\ref{eq:pvamp}), 
from both strong interactions and 
higher dimension weak interactions\cite{KSa}.
Therefore, we do not pursue this calculation beyond leading order.

In the energy regime where $\omega\ll m_\pi$, the 
dominant part of the parity-violating amplitude
in eq.~(\ref{eq:pvamp})
is reproduced by a lagrange density of the form,
\begin{eqnarray}
{\cal L}^{(pv)}_{\gamma\gamma} & = & 
\overline{N}
\left( W_{\gamma\gamma}^{(0)}\ +\  W_{\gamma\gamma}^{(1)}\
  \tau^3\right)\ v_\mu  S_\nu N\ 
F^{\mu\alpha} F_\alpha^{\ \nu}
\ \ \ .
\label{eq:pvgg}
\end{eqnarray}
where $W_{\gamma\gamma}^{(0),(1)}$ are the isoscalar and isovector 
dimension seven coupling constants 
\begin{eqnarray}
W_{\gamma\gamma}^{(0)} & = & 0\ \ ,\ \ 
W_{\gamma\gamma}^{(1)} \ =\ 
{e^2 g_A h_{\pi NN}^{(1)}\over 12\pi^2 f m_\pi^2}
\ \ \ .
\label{eq:match}
\end{eqnarray}

The differential cross section for $\gamma \overrightarrow p$
Compton scattering resulting from the amplitudes in
eqs.~(\ref{eq:ampst}) and (\ref{eq:pvamp}) is, to leading order in the 
chiral expansion and weak interaction,
\begin{eqnarray}
& & {d\sigma\over d\Omega} \ =\ 
{\alpha^2\over 2 M_N^2} 
 \left[ 1+\cos^2\theta
\right.\nonumber\\
& & \left.
\ \ 
+\ \eta {g_A h_{\pi NN}^{(1)} M_N\over 12\pi^2 f m_\pi^2}\ \omega^2\ 
 (1+\cos\theta)\ 
{\rm Re} \left[\ 
\tilde {\cal F}_1(\omega,\theta) 
\ (1+\cos^2\theta)
\ +\ 
\tilde {\cal F}_2(\omega,\theta)
\ \cos\theta (1-\cos\theta)
\right.\right.\nonumber\\
& & \left.\left.
\qquad\qquad\qquad\qquad\qquad\qquad \ -\ 
\tilde {\cal F}_3(\omega,\theta)\ 
{\omega^2\over 15 m_\pi^2}\cos\theta (1-\cos^2\theta)
\right]\ 
\right]
\ ,
\label{eq:crossfull}
\end{eqnarray}
where $\eta=+1 (-1)$ for the proton spin-polarized parallel (anti-parallel)
to the direction of the incident photon.
The functions $\tilde {\cal F}_i$ are 
\begin{eqnarray}
\tilde {\cal F}_1(\omega,\theta) & = & -{6 m_\pi^2\over \omega} 
{\cal F}_1 (\omega,\theta)
\ \rightarrow\ 
1 \ +\ {\omega^2\over 15 m_\pi^2}\left(3+\cos\theta\right)
\ +\ ...
\nonumber\\
\tilde {\cal F}_2(\omega,\theta) & = & -{6 m_\pi^2\over \omega} 
{\cal F}_2 (\omega,\theta)
\ \rightarrow\ 
1 \ +\ {2 \omega^2\over 15 m_\pi^2}\cos\theta
\ +\ ...
\nonumber\\
\tilde {\cal F}_3(\omega,\theta) & = &   {90 m_\pi^4\over \omega} 
{\cal F}_3 (\omega,\theta)
\ \rightarrow\ 1 \ +\ {2\omega^2\over 7 m_\pi^2}\cos\theta
\ +\ ...
\ \ \ ,
\label{eq:fdef}
\end{eqnarray}
and have been normalized such that 
$\tilde {\cal F}_i (\omega\rightarrow 0,\theta)\rightarrow 1$
and are slowly varying functions of $\omega$.

The parity-violating asymmetry defined by the difference in cross section for
$\eta=+1$ and $\eta=-1$ normalized to the sum, is
\begin{eqnarray}
& & A^{\gamma\gamma}(\omega,\theta) \ =\  
{g_A h_{\pi NN}^{(1)} M_N \omega^2\over 12\pi^2 f m_\pi^2}
\ {1+\cos\theta\over 1+\cos^2\theta}
\nonumber\\
& & {\rm Re}\left[\ 
\tilde {\cal F}_1(\omega,\theta)
\ (1+\cos^2\theta)
\ +\ 
\tilde {\cal F}_2(\omega,\theta)
\ \cos\theta (1-\cos\theta)
 \ -\ 
\tilde {\cal F}_3(\omega,\theta) 
{\omega^2\over 15 m_\pi^2}\cos\theta (1-\cos^2\theta)
\right]
\ \ \ .
\label{eq:asymmfull}
\end{eqnarray}

For a numerical estimate of the magnitude of the asymmetry,
we consider forward scattering, $\theta=0$,
where 
\begin{eqnarray}
A^{\gamma\gamma}(\omega,0) & = & 
{g_A h_{\pi NN}^{(1)} M_N \omega^2\over
6\pi^2 f m_\pi^2} \tilde {\cal F}_1 (\omega,0)
\ =\ 
1.5\times 10^{-9}
\left({h_{\pi NN}^{(1)}\over 5\times 10^{-7}}\right)
\left({\omega\over 20\ {\rm MeV}}\right)^2
\tilde {\cal F}_1 (\omega,0)
\ \ \ .
\label{eq:aszero}
\end{eqnarray}
The asymmetry  for  $100\  {\rm MeV}$ photons is
$A^{\gamma\gamma}(100,0)\sim 5\times 10^{-8}$
assuming the naive value for  $ h_{\pi NN}^{(1)}$, which is
comparable to the forward-backward asymmetry expected in
$\overrightarrow n+p\rightarrow d+\gamma$.

In conclusion, we have computed the leading contribution to 
parity-violation in $\gamma\overrightarrow N$ scattering.
It arises from one-loop pion graphs with one insertion of the 
parity-violating pion-nucleon interaction described by $h_{\pi NN}^{(1)}$,
and scales like $1/m_\pi^2$ in the chiral limit.
The absence of a $\gamma\gamma Z^0$ interaction means that a
measurement of this asymmetry will provide a background free
determination of  $h_{\pi NN}^{(1)}$, up to corrections suppressed by 
$m_\pi/\Lambda_\chi$ and $\omega/\Lambda_\chi$, i.e. $\sim 15\%$ 
for photon energies below
the pion photo-production threshold.
While this asymmetry, along with all other parity-violating asymmetries in the
few-nucleon sector, is small $\sim 10^{-8}$, the high intensity photon 
sources that are currently in operation (such as the FEL at Duke), 
or may come on-line in the future, provide
hope that this asymmetry can be measured.

\vskip 1in

This work is supported in part by the U.S. Dept. of Energy under
Grants No. DE-FG03-97ER4014 and DOE-ER-40561.

\vfill\eject

\end{document}